\documentstyle[12pt]{article}
\input epsf
\def\be{\begin{equation}}
\def\ee{\end{equation}}
\def\l{\label}
\def\refe#1{(\ref{#1})}

\def\ms{$M_{\scriptscriptstyle String}$}

\def\LEP2{{LEPII}}
\def\mg{$m_{3/2}$}
\def\th{$\theta$}

\def\npb#1#2#3{    {\it Nucl. Phys. }{\bf B #1} (19#2) #3}

\def\plb#1#2#3{    {\it Phys. Lett. }{\bf B #1} (19#2) #3}
\def\prd#1#2#3{    {\it Phys. Rev. }{\bf D #1} (19#2) #3}
\def\prep#1#2#3{   {\it Phys. Rep. }{\bf #1} (19#2) #3}
\def\prl#1#2#3{    {\it Phys. Rev. Lett. }{\bf #1} (19#2) #3}

\def\zpc#1#2#3{    {\it Zeit. f\"ur Physik }{\bf C #1} (19#2) #3}

\def\ibid#1#2#3{   {\it ibid. }{\bf #1} (19#2) #3}

\begin{document}
\begin{titlepage}
\baselineskip=14pt
\begin{center}
\rightline{BA/97/10}
\vskip1.5truecm

{\Large \bf From $b \rightarrow s \gamma$ to the LSP Detection
Rates in Minimal String Unification Models}\\ 
\vspace{3ex}
{\bf Shaaban Khalil}$^{a,b}$, {\bf Antonio Masiero}$^{c}$ {\bf and Qaisar 
Shafi}$^{d}$\\
{\it
\vspace{1ex} a) International Center For Theoretical Physics,\\
 ICTP, Trieste, Italy.}\\
{\it
\vspace{1ex} b) Ain Shams University, Faculty of Science\\
 Department of Mathematics, Cairo, Egypt.}\\
{\it
\vspace{1ex} c)International School for Advanced Studies,\\
SISSA, via Beirut 2-4, I-34100 Trieste, Italy.}\\
{\it
\vspace{1ex} d)Bartol Research Institute, University of Delaware\\ 
Newark, DE 19716, USA}\\

\vspace{6ex}
{ABSTRACT}
\end{center}
\begin{quotation}
We exploit the measured branching ratio for
$b \rightarrow  s\gamma$ to derive lower limits on the sparticle and
Higgs masses in the minimal string unification models. For the
LSP('bino'), chargino and the lightest Higgs, these turn out to be 50, 90
and 75 GeV respectively.
Taking account of the upper bounds on the mass spectrum from the LSP
relic abundance, we estimate the direct detection rate for the latter to
vary from $10^{-1}$ to $10^{-4}$ events/kg/day. The muon flux, produced
by neutrinos from the annihilating LSP's, varies in the range $10^{-2} -
10^{-9}$
muons/m$^2$/day.
\end{quotation} \end{titlepage} 
\vfill\eject
\section{Introduction}
\hspace{0.75cm}The supersymmetric theories provide the most promising 
framework to extend the standard model (SM) \cite{nilles}. Local 
supersymmetry (SUSY) breaking induces soft SUSY breaking terms such as 
gaugino masses, scalar masses and trilinear and bilinear couplings of 
scalar fields in global SUSY models. The values of these soft terms 
determine the phenomenological properties of the models. Four-dimensional 
superstrings is a promising candidate for the unification of all 
interactions, and so far the only candidate theory for
a consistent treatment of gravity on the quantum level. In  
four-dimensional string approaches such as the Calabi-Yau~\cite{calabi} or
orbifold models \cite{orbifold}, the expressions for
the K\"ahler and the gauge kinetic functions of supergravity are known. In 
recent papers \cite{ibanez1},\cite{kap},\cite{munoz} 
the soft SUSY breaking terms have been derived from superstring theories 
under the assumption that SUSY is broken by the vacuum expectation value
(vev) of the F-terms of the dilaton field  $S$ and/or the moduli fields
$T_i$. These gauge singlet fields are  generically present in a large
class of four-dimensional 
models, and their coupling to the gauge non-singlet matter are suppressed
by powers of the Planck mass. This makes them `natural' candidates to 
constitute the SUSY-breaking ``hidden sector" which is needed
in many phenomenological 
models of low-energy SUSY. The vev of the real part of $S$ gives the 
inverse square of the tree-level gauge coupling constant, and the vevs 
of the moduli fields $T_i$ parameterize the size and shape of the 
compactified space.  In Ref.~\cite{ibanez2} it 
was assumed that only the dilaton field and an overall modulus field $T$ 
contribute to the SUSY breaking, and the soft SUSY breaking 
terms were parameterized by a goldstino angle \th , the gravitino mass 
\mg\ and a set of parameters known as modular weights. The case with 
multi-moduli fields is discussed in Ref.\cite{kobay}.\\

	We choose the modular  weights such that one can have
the appropriate large string threshold  corrections to ensure the
`merging' of the three `low energy' gauge couplings at the GUT scale. This
is the so-called minimal string
unification model~\cite{ross}. It is worth repeating here that the
natural unification scale in superstring models is $M_{string}\sim
0.5\times g_{string} \times 10^{18}$ GeV, where $g_{string}=(Re
S)^{-1/2}\sim 0.7$.
However, the merging of the gauge coupling constants with the particle
content of the MSSM (minimal SUSY SM) takes place at a scale $M_X \sim
3\times 10^{16}$ GeV. Several mechanisms have been proposed in order to
explain this $M_X-M_{string}$ discrepancy. The string threshold
corrections is an elegant possibility since it does not require any new
particles beyond the MSSM ones. \\

	The investigation of the phenomenology  of this approach leads to a  
number of `low energy' predictions which can be tested at \LEP2, Tevatron
and LHC, as we have emphasized in Refs.\cite{vissani} and \cite{shafi}.
We have shown  that the lightest neutralino
in this model also happens to be the lightest supersymmetric particle
(LSP), and it is almost a `pure' bino where the gaugino function is
larger than 0.97. Moreover, the cosmic 
abundance of the LSP puts important constraints on the underlying 
supersymmetry breaking parameters, namely one obtains an upper bound on the 
gravitino mass of about 600 GeV, which leads to upper bounds on the sparticle
spectrum of the model.\\

	In this paper we wish to focus on the reduction of the allowed 
parameter space that follows from the $b \rightarrow s \gamma $ physics.  
(It is well known that the constraints from $b \rightarrow s \gamma $
cause a dramatic reduction of the allowed MSSM parameter space) In models
with dilaton/moduli SUSY breaking, we find that the $b
\rightarrow s \gamma$ constraint imposes a lower bound on the gravitino
mass which implies stringent lower bounds on the entire SUSY spectrum (for
instance, the lower bound that we obtain on the chargino mass exceeds
by a small amount its present experimental bound of 85 GeV). We provide
estimates for the detection rates in direct and indirect neutralino 
searches within the range of the parameter space which satisfies all the 
particle accelerator and relic abundance constraints.\\

	This paper is organized as follows. In section 2 we review the 
soft SUSY breaking terms obtained in Ref.\cite{kap} and their 
parameterization following Ref.~\cite{ibanez2}. Also, we study the 
effect of leaving free the $B$-parameter. Section 3 deals with the constraints
on SUSY parameter space from the $ b 
\rightarrow s \gamma$ decay. In section 4, we study the direct and indirect
LSP detection rates in the allowed range of the parameter space. We
show that the event rate $R$ lies between 0.1 and $10^{-4}$
events/kg/day while
the muonic flux $\Gamma$ satisfies $10^{-9} \leq \Gamma \leq
10^{-2}$ muon/m$^2$/day. We give our conclusions in section 5.

\section{Minimal String Unification}
\hspace{0.75cm}In this section we give a brief review of the construction 
of the soft SUSY breaking terms in the minimal string 
unification scheme~\cite{ross} and their low energy  
implications~\cite{vissani},~\cite{shafi}. A supergravity lagrangian is
characterized by the K\"ahler potential 
K, a superpotential $W$ and the gauge function $f_a$, where $a$ refers to the 
gauge group. In the case of orbifold four-dimensional superstrings, the 
K\"ahler potential has the form \cite{witten}
\be
K= - \log (S+S^*) -3 \log (T+T^*) + \sum_i (T+T^*)^{n_i} \phi_i \phi_i^*.
\ee
Here $S$ is the dilaton field which couples universally in all string 
models, $T$ is the overall modulus whose real part gives the 
volume of  the compactified space, and $\phi_i$ are the chiral superfields. 
The $n_i$ are integers, called 
the modular  weights of matter fields. At tree level the gauge kinetic 
function is given by $f_a=k_a S$, where $k_{a}$ is the 
Kac-Moody level of the gauge factor. In the
phenomenological analysis that follows, 
$k_{3}=k_{2}=\frac{3}{5}k_{1}=1$. 
Assume that the fields which contribute to SUSY breaking are $S$ and $T$
through a non 
vanishing vevs of their auxiliary fields $F^S$ and $F^T$ respectively. We
can take the following parameterization for the vevs:
\be
(G_{\bar{S}S})^{1/2} F^S = \sqrt{3} m_{3/2} \sin\theta,
\ee
\be
(G_{\bar{T}T})^{1/2} F^T = \sqrt{3} m_{3/2} \cos\theta.
\ee
Here $G_{ij}=\frac{\partial^2 G}{\partial \phi_i \partial \phi_j}$, 
$G=-3\ln(\frac{K}{3}) + \ln \vert W \vert^2$ and $e^{G/2}=m_{3/2}^2$ is the
gravitino mass. The angle $\theta$ 
parameterizes the direction of the goldstino field $\tilde{\eta}$ (the
goldstino fermion is eaten by the gravitino in the process of SUSY breaking)
in the $S,T$ field space:
$$\tilde{\eta} = \sin\theta \tilde{S} +\cos\theta \tilde{T},$$
where $\tilde{S}$ and $\tilde{T}$  are the canonically normalized
fermionic partners of the scalar fields S and T. Thus this angle is called the
goldstino angle. 

The soft breaking terms have the form 
\be 
m_i^{2}=m_{3/2}^{2}(1+n_i \cos^{2}{\theta}),
\l{scalar1}
\ee
\be
M_{a}=\sqrt{3}m_{3/2}\frac{k_{a}ReS}{Ref_{a}}\sin{\theta}
+m_{3/2}\cos{\theta}\frac{B_{a}^{'}(T+T^{\ast})\hat{G}_{2}(T,T^{\ast})}
{32{\pi}^{3}Re{f}_{a}},
\l{gaugino1}
\ee
\be
A_{ijk}=-\sqrt{3}m_{3/2}\sin{\theta}-
m_{3/2}\cos{\theta}(3+n_i+n_j+n_k),
\l{trilinear1}
\ee
\be
B_{\mu}= m_{3/2}\left [-1 -\sqrt{3} \sin{\theta} 
-\cos{\theta}(3+n_{H_1}+n_{H_2}) \right], 
\l{bilinear1}
\ee
where the definitions of the $B^{'}_{a}$, $\hat{G}_{2}$ functions are
given in Ref.\cite{ibanez2}. It is clear that if $S$ contributes dominantly to SUSY 
breaking ($\theta=\pi/2$), we obtain the well known universal soft scalar
and gaugino masses. Otherwise, the soft scalar masses depend on their
modular weights and $T$-dependent threshold corrections that lead to non
universal gaugino masses.\\

	In Ref.\cite{ibanez2} two sources for the B-parameter were 
considered, labeled by $B_Z$ and $B_{\mu}$. Here $B_{\mu}$ is the 
coefficient of the $H_1-H_2$ mixing term in the scalar Higgs sector
potential. The source of $B_Z$ is the mechanism of Ref.\cite{masiero1}, 
where $\mu$ arises only from couplings in the K\"ahler potential. The 
source of $B_{\mu}$ is considered in Ref.\cite{ibanez2}, with $\mu$ 
coming solely from the $S$ and $T$ sector. In general, there could be an 
admixture of these two cases. In our previous analysis we have 
focused on just $B_{\mu}$, since this 
option for $B$ allows a larger region of SUSY parameter space for
electroweak radiative breaking. Here we will study the effect of 
leaving $B$ as a free parameter to be 
determined from the electroweak breaking conditions.\\

	An interesting input in order to further 
constrain the form of the soft terms is the requirement of
the gauge couplings unification. As mentioned in the introduction, 
in string unification scenarios, 
the gauge coupling constants $g_a$ of the three SM interactions are 
related at the string scale $M_{String}$ by
\be
k_3 g_3^2 =k_2 g_2^2 =k_1 g_1^2=\frac{8\pi}{\alpha'}
G_{Newton}=g^2_{String}.
\ee
In fact, it has been shown~\cite{kap2} that the string scale is of order 
$g_{String} \times 0.5 \times 10^{17}$ GeV. Thus one finds that there is 
an order of magnitude discrepancy between the SUSY GUT and the string
unification scales. Here we assume the so-called minimal 
string unification~\cite{ross}, namely that the only `light' particles
are the standard MSSM ones. We rely on the string threshold 
contributions to `cover' the gap between the unification and the 
string scales. It has been shown~\cite{ross} that 
the following values of the modular weights lead to 
good agreement for $\sin^2 \theta_W$ and $\alpha_3$.
\be
n_{Q_L}=n_{D_R}=-1,\hspace{0.5cm} n_{u_R}=n_{H_1}=-2,
\hspace{0.5cm} n_{L_L}=n_{E_R}=n_{H_2}=-3.
\l{modular}
\ee
The above values of the modular weights are interesting because they
provide  an explicit model with non-universality which has 
phenomenological implications which differ from the case of the 
universal boundary conditions. 
From eq.\refe{scalar1} and
above set \refe{modular} of values of the modular weights, we readily
obtain the value of the scalar masses. Notice that if one require the
absence of tachyonic mass at the compactification scale then we get 
$\cos^2\theta \leq 1/3$. The asympotatic gaugino masses read:
\begin{eqnarray}
M_{3}&=&\sqrt{3}m_{3/2}(\sin{\theta}+0.12 \cos{\theta}),
\nonumber
\l{gaugino2}\\
M_{2}&=&\sqrt{3}m_{3/2}(\sin{\theta}+0.06 \cos{\theta}),
\\
M_{1}&=&\sqrt{3}m_{3/2}(\sin{\theta}- 0.02 \cos{\theta}).
\nonumber
\end{eqnarray}
The trilinear scalar coupling  $A_t$ which is related to the top quark 
Yukawa coupling is given by
\be
A_{t}=-m_{3/2}(\sqrt{3}\sin{\theta}-3 \cos{\theta}).
\l{trilinear2}
\ee
Finally $B_{\mu}$ has the form
\be
B_{\mu}=m_{3/2}(-1-\sqrt{3}\sin{\theta}+2 \cos{\theta}).              
\l{bilinear2}
\ee
Given the boundary conditions in eqs.
(\ref{gaugino2}-\ref{bilinear2}) 
at the compactification scale, we determine the evolution of the
various couplings according to their renormalization group
equation (RGE) to finally compute the mass spectrum of the SUSY particles
at the weak scale. The electroweak symmetry breaking requires the 
following conditions among the renormalized quantities $\mu_1^2$,
$\mu_2^2$ and $\mu_3^2$: \\ 
\be
\mu_1^2 +\mu_2^2 > 2\mu_3^2, \hspace{1cm} \vert \mu_3 \vert ^4 > \mu_1^2
\mu_2^2,
\ee
and
\be
\mu^2 = \frac{ m_{H_1}^2 -m_{H_2}^2 \tan^2\beta}{\tan^2\beta - 1}
- \frac{M_Z^2}{2},
\hspace{1.5cm} \sin 2\beta= \frac{-2  B \mu }{m_{H_1}^2+m_{H_2}^2+ 2\mu^2}.
\l{minimization1}
\ee
Here $\tan\beta= {\langle H_2^0 \rangle}/{\langle H_1^0 \rangle}$ is the 
ratio of the two Higgs vevs that give masses to the up and down type 
quarks, and $m_{H_1}$ , $m_{H_2}$ are the Higgs masses at the electroweak 
scale. $\mu_1^2$, $\mu_2^2$ and $\mu_3^2$ satisfying the boundary
conditions at \ms:
\be
\begin{array}{l}
\mu_i^2= m_{H_i}^2 + \mu^2  \hspace{2cm}i=1,2\\
\mu_3^2=-B\mu.
\end{array}
\ee
Using equations \refe{minimization1} we find that 
$\mu$ and $\tan\beta$ are specified in terms of the goldstino angle \th\ 
and the gravitino mass \mg. It turns out that if only the top 
Yukawa coupling is of order unity, $\tan\beta$ is close to 2 and $\vert \mu 
\vert$ is quite large. For instance, $\mu$ is about 350 GeV for the lower 
chargino bound of 84 GeV. 

As explained in Ref.\cite{vissani} a further constraint on the parameter
space is entailed 
by the demand  of color and electric charge conservations. In particular, 
with the latter constraint one finds that 0.98 rad $\leq \theta \leq $2  
rad. In the case with $B$ as a free parameter, we can
determine it from equation \refe{minimization1} in terms of $\tan \beta$.
Fig.1 shows  the ratio of $B$ ( at compactification scale ) to $m_{3/2}$
versus the gravitino mass, for $\tan \beta=6$. We note that
the sign of $B$ in general is opposite that of $\mu$ for the correct
electroweak symmetry breaking. For the value of $m_{3/2}$ fixed different
values of $B/ m_{3/2}$ in this figure corresponds to different
values of $\theta$. In the same way, all the figures are plotted
corresponding to different values of $\theta$.

\vspace{0.5cm}
\begin{center}
\input rat1.tex
\end{center}
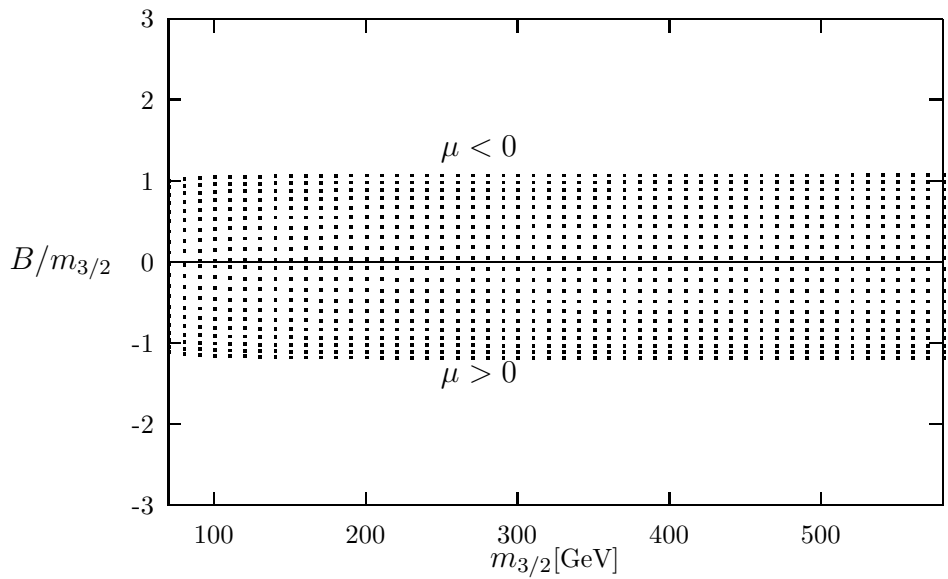
Figure 1: The ratio of $B$ (at compactification scale) to 
$m_{3/2}$ versus the gravitino mass, with $\tan\beta=6$.
\vspace{0.5cm}
  
	We have previously shown~\cite{shafi} that the lightest neutralino
is almost a `pure' bino and is predicted to be the lightest sparticle 
(LSP). The LSP is a leading cold dark matter candidate and expected to
play an important role in the large scale structure formation.
Assuming a relic density parameter $0.1 \leq \Omega_{LSP} \leq 
0.9$,  with the Hubble parameter $h$ in the range $ 0.4 \leq h \leq 0.8 $,
we found that the maximum value of neutralino relic density
$\Omega_{LSP} h^2$ imposes an upper bound on \mg\ which is 
very sensitive to \th.  In turn, this leads to a stringent upper bound on 
the LSP mass of about 160 GeV in the case of pure dilaton SUSY breaking,
while this bound reaches 300 GeV in the case of $\theta=0.98$ rad which
represents the maximum moduli contribution to SUSY breaking in this model.
There is no point in the parameter space (\mg, \th), after imposing
the experimental constraints on the sparticles, that leads to
$\Omega_{LSP}h^2$ less than the minimal value (0.014). Hence requiring the
LSP to provide the correct amount of the cold dark matter does not lead to
a lower bound on \mg. However, in the next section we will  
show that the recent observational bounds on $b\rightarrow s
\gamma $ impose lower bounds on the sparticle masses of this model.
Consequently, we are able to provide both the lower and upper bounds on
the sparticle spectrum.

\section{\large{\bf{Constraints from $b\rightarrow s \gamma $}}}
\hspace{0.75cm} In this section we focus on the constraints on the 
parameter space (\mg,\th) which come from $b \rightarrow s \gamma$ 
decay. The recent observation of this process by the CLEO collaboration
\cite{amer}, $1\times 10^{-4} < BR(b \rightarrow s \gamma) < 4 \times
10^{-4}$ (at $95\%\ c.l.$), has stirred interest in the possible
bounds obtainable for supersymmetric models because of new contribution to
this process~\cite{masiero2},\cite{bertolini}. For a recent discussion in
the context of the effective supergravities from string theories see  
Ref.\cite{karniotis}.\\

      In MSSM there are additional contributions to the decay besides the
SM diagrams with a W-gauge boson and an up-quark in the loop. The new
particles running in the loop are: charged Higgs
$(H^{\pm})$ and up-quark, charginos $(\chi^-)$ and up-squarks, gluino and
down-squarks, neutralinos and down-squarks \cite{masiero2}. The total 
amplitude for the decay is the sum of all these contributions. The 
inclusive branching ratio for $b \rightarrow s \gamma$ normalized to 
$BR(b\rightarrow c e\bar{\nu})$ is given by ~\cite{masiero2}
\be
 \frac{BR(b \rightarrow s \gamma)}{BR(b \rightarrow c e\bar{\nu})}=
\frac{\mid V_{ts}^* V_{tb}\mid^2}{\mid V_{cb} \mid^2} \frac{6
\alpha_{em}}{\pi} \frac{[\eta^{16/23} A_{\gamma} + \frac{8}{3} (\eta^{14/23}
-\eta^{16/23}) A_g + C]^2}{I(x_{cb})[1-\frac{2}{3\pi} \alpha_S(m_b)
f(x_{cb})]}.
\ee
Here $\eta = \frac{\alpha_S(m_W)}{\alpha_S(m_b)}$, and $C$ represents the
leading-order QCD 
corrections to $b \rightarrow s \gamma$ amplitude at the scale $Q=m_b$ 
\cite{misiak}. The function $I(x)$ is given by
$$I(x)=1-8 x^2 +8 x^6 -x^8 -24 x^4 \ln x, $$
and $x_{cb} = \frac{m_c}{m_b}$, while $ f(x_{cb})=2.41$ is a QCD
correction factor. The amplitude $A_{\gamma}$ is from
the photon penguin vertex, and the $A_g$ from the gluon
penguin vertex. The relevant contributions which we will consider 
come from the SM diagram plus those with the top quark and charged Higgs,
and up-squarks and charginos running in the loop. Following the notation
of 
Ref.\cite{guduice} these contributions read:
\begin{eqnarray}
A^{SM}_{\gamma,g} & = &\frac{3}{2}\frac{m_{t}^{2}}{M^{2}_{W}}
f^{(1)}_{\gamma,g}\left(\frac{m^{2}_{t}}{M^{2}_{W}}\right), \nonumber \\
A_{\gamma,g}^{H^{-}}&=&\frac{1}{2}\frac{m^{2}_{t}}{m^{2}_{H}}   
\left[\frac{1}{\tan^{2}\beta}f_{\gamma,g}^{(1)}\left(
\frac{m^{2}_{t}}{m^{2}_{H}}\right)+f_{\gamma,g}^{(2)}\left(
\frac{m^{2}_{t}}{m^{2}_{H}}\right)\right], 
\l{ampl1}\\
A_{\gamma,g}^{\chi^{-}} &   =   &A_{\gamma,g}^{\chi^{-}})_1 
+A_{\gamma,g}^{\chi^{-}})_2
+A_{\gamma,g}^{\chi^{-}})_3+A_{\gamma,g}^{\chi^{-}})_4,  
\nonumber
\end{eqnarray}
where $A_{\gamma,g}^{\chi^{-}})_i$ are given by
\begin{eqnarray}
A_{\gamma,g}^{\chi^{-}})_1 &   =   & \sum_{j=1}^{2} \frac{M^{2}_W}
{M^{2}_{\chi_j}}|V_{j1}|^{2}f_{\gamma,g}^{(1)}\left(\frac{m^{2}_{\tilde{c}}}
{M^{2}_{\chi_j}}\right),
\nonumber\\
A_{\gamma,g}^{\chi^{-}})_2 &   =   &-  \sum_{j,k=1}^{2} 
\frac{M^{2}_W}{M^{2}_{\chi_j}}
\left|V_{j1}T_{k1}-\frac{V_{j2}m_tT_{k2}}{\sqrt{2}
M_W\sin\beta}\right|^{2} f_{\gamma,g}^{(1)}\left(\frac{m^{2}_{\tilde
{t}_k}}{M^{2}_{\chi_j}}\right),
\nonumber\\
A_{\gamma,g}^{\chi^{-}})_3 &   =   & - \sum_{j=1}^{2} 
\frac{M_W}{M_{\chi_j}} \frac{U_{j2}V_{j1}}{\sqrt{2}\cos\beta} 
f_{\gamma,g}^{(3)}\left(\frac {m^{2}_{\tilde{c}}}{M^{2}_{\chi_j}}\right), 
\label{apml2}\\
A_{\gamma,g}^{\chi^{-}})_4 &   =   & \sum_{j,k=1}^{2} 
\frac{M_W}{M_{\chi_j}} 
\frac{U_{j2}}{\sqrt{2}\cos\beta}                        
\left( V_{j1}T_{k1}-V_{j2}T_{k2}\frac{m_t}{\sqrt{2}M_{W}\sin\beta}\right)
T_{k1}f_{\gamma,g}^{(3)}\left(\frac{m^{2}_{\tilde{t}_k}}{M^{2}_{\chi_j}}
\right).
\nonumber
\end{eqnarray}
The functions $f_{\gamma,g}^{(i)}$, $i=1,2,3$ are given 
in \cite{masiero2}, $V$ and $U$ are the unitary matrices which
diagonalise the chargino mass matrix, while $T$ diagonalises the
stop mass matrix\footnote{We use the sign convention of $\mu$ in the 
chargino mass matrix opposite to that adopted in the Haber and Kane 
report~\cite{nilles}}. \\

	As it is known, the charged Higgs contribution always interferes
constructively with the SM contribution. The chargino contribution could 
give rise to a substantial destructive interference with the SM and Higgs 
amplitudes, depending on the sign of $\mu$, the value of $\tan\beta$ and 
the mass splitting between the stop masses. This way of presenting the 
chargino amplitude in equation \refe{ampl1} by splitting it into four
pieces can help us to show when the chargino contribution can
significantly reduce $b\rightarrow s \gamma $~\cite{ng}. It turns out
numerically that the magnitudes of $A_{\gamma,g}^{\chi^{-}})_1$ and
$A_{\gamma,g}^{\chi^{-}})_2$ are less than those of
$A_{\gamma,g}^{\chi^{-}})_3$ and $A_{\gamma,g}^{\chi^{-}})_4$, especially 
with the \LEP2 lower bound on the chargino mass $m_{\chi} > 84$ GeV. 
Moreover, we can observe that the sum of 
$A_{\gamma,g}^{\chi^{-}})_3+A_{\gamma,g}^{\chi^{-}})_4$ is identically 
zero in the limit of degenerate up-squark masses.
On the other hand, $A_{\gamma,g}^{\chi^{-}})_3$ and 
$A_{\gamma,g}^{\chi^{-}})_4$ each can be quite large because they are
enhanced by large $\tan\beta$. \\

First we make our analysis in the context of the choice $B_{\mu}$
(eq.\refe{bilinear2}) for the $B$ parameter. Hence $\tan \beta$ is fixed 
to be $\simeq 2$. In Figs.2 and 3 we show the values of the 
$BR(b\rightarrow s \gamma )$  corresponding to the gravitino mass \mg\ in 
the allowed range we have  determined in Ref~\cite{shafi} for $\mu <0$ 
and $\mu>0$ respectively.
\vspace{0.5cm}
\begin{center}
\input bs1.tex
\end{center}
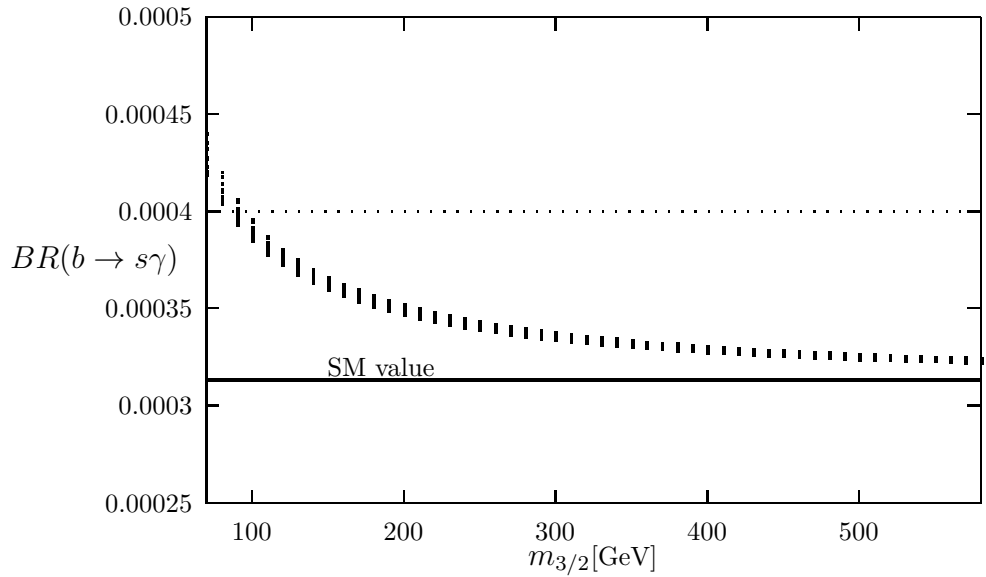
Figure 2: The branching ratio $ BR(b\rightarrow s \gamma)$ versus
$m_{3/2}$ with $\mu<0$, while $\tan \beta \simeq 2$ from electroweak
breaking.

\vspace{0.5cm} 

\begin{center}
\input bs2.tex
\end{center}
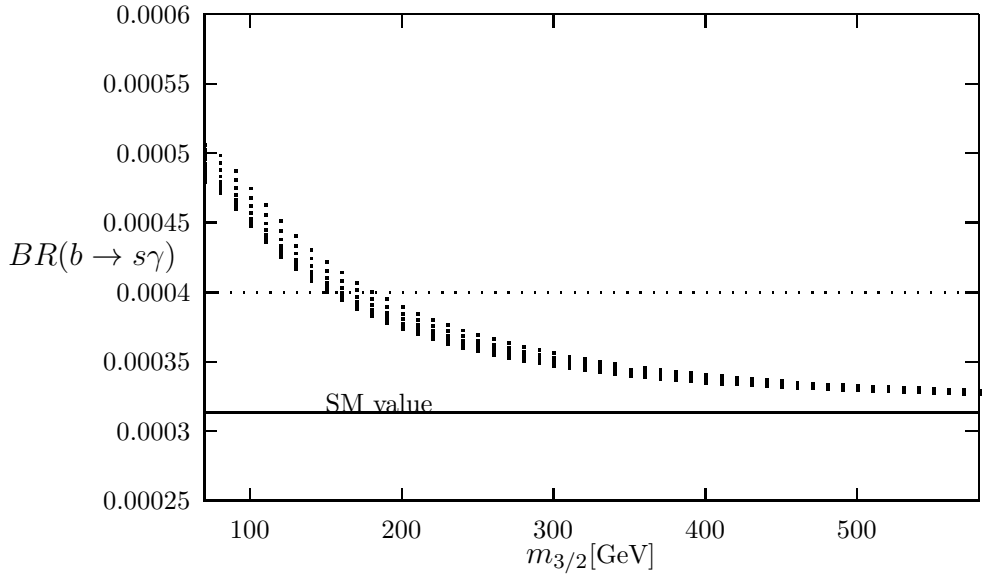
Figure 3: The branching ratio $BR(b\rightarrow s \gamma)$ versus
$m_{3/2}$, with $\mu>0$ and $\tan \beta \simeq 2 $.

\vspace{0.5cm}

It is remarkable that for $\mu <0$, even taking the experimental upper
bound $BR(b\rightarrow s \gamma)<4\times 10^{-4}$ 
we obtain a lower bound of $\simeq 90$ GeV 
on the gravitino mass while, for $\mu >0$ the lower bound is 
150 GeV. This can be explained as follows. For $\mu <0$, the
chargino contribution gives a destructive interference with the SM and
$H^+$ contributions, but it is smaller in magnitude.
This is due to the fact that $\tan \beta$ is of order 2, and 
the splitting of the two stop mass eigenstates $m_{\tilde{t}_{1,2}}^2$ is
small since the L-R entry in the stop mass matrix is quite
small with respect to the value of the diagonal elements which get a
large gluino contribution in the renormalization group evolution.
For $\mu>0$ the chargino gives a constructive interference with the SM
and $H^+$, and this makes the branching ratio  
larger than the experimental upper bound, unless the Higgs and chargino 
masses are sufficiently large.\\

Now we relax the assumption $B=B_{\mu}$ which forces $\tan\beta\simeq 2$
and let $B$ be a free parameter. For larger values of $\tan \beta$ and
$\mu<0$ we expect the
chargino contribution to give rise to substantial destructive 
interference and the branching ratio of $b\rightarrow s \gamma$ 
can be less than the standard model value as shown in Fig.4.\\
\vspace{0.5cm}
\begin{center}
\input bs3.tex
\end{center}
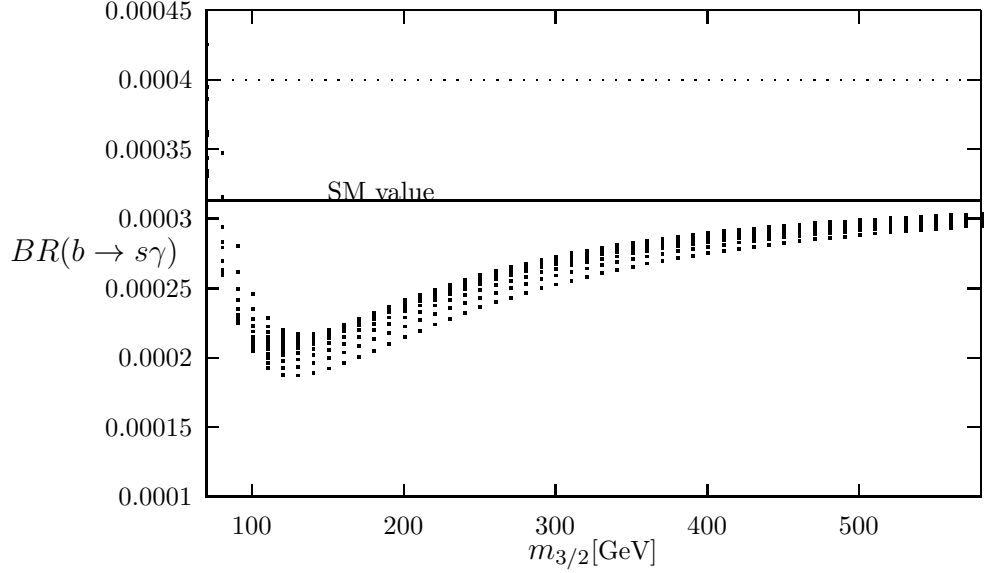
Figure 4: The $b\rightarrow s \gamma$ branching ratio versus $m_{3/2}$,
with $\mu<0$ and $\tan \beta=20$.
\vspace{0.5cm}

          The new constraints from $b\rightarrow s \gamma$ shrink
the allowed parameter space of the model as shown in Fig.5 for the case
$\mu <0$. The model predicts that the mass of the chargino is greater than
90 GeV. The lower bound in the right selectron 
mass turns out to be 65 (110) GeV 
if $\theta=0.98 (\pi/2)$. Actually, the discovery of a right
selectron  with  mass less than the chargino mass would be signal for a
departure from the (pure dilaton dominated) universal soft SUSY
breaking scenario. Also of much interest are the  lower bounds  on the
lightest neutralino and lightest Higgs: they are 50 GeV and 75 GeV
respectively. As we said, all these bounds apply in the case $B=B_{\mu}$.
On the other hand, if we let $B$ free then for $\mu <0$ no lower bound on
\mg\ is obtained.
\vspace{0.5cm} \begin{center}
\pagestyle{empty}
\setlength{\unitlength}{0.240900pt}
\ifx\plotpoint\undefined\newsavebox{\plotpoint}\fi
\sbox{\plotpoint}{\rule[-0.200pt]{0.400pt}{0.400pt}}%
\begin{picture}(1500,900)(0,0)
\font\gnuplot=cmr10 at 10pt
\gnuplot
\sbox{\plotpoint}{\rule[-0.200pt]{0.400pt}{0.400pt}}%
\put(220.0,113.0){\rule[-0.200pt]{292.934pt}{0.400pt}}
\put(220.0,113.0){\rule[-0.200pt]{4.818pt}{0.400pt}}
\put(198,113){\makebox(0,0)[r]{0}}
\put(1416.0,113.0){\rule[-0.200pt]{4.818pt}{0.400pt}}
\put(220.0,222.0){\rule[-0.200pt]{4.818pt}{0.400pt}}
\put(198,222){\makebox(0,0)[r]{100}}
\put(1416.0,222.0){\rule[-0.200pt]{4.818pt}{0.400pt}}
\put(220.0,331.0){\rule[-0.200pt]{4.818pt}{0.400pt}}
\put(198,331){\makebox(0,0)[r]{200}}
\put(1416.0,331.0){\rule[-0.200pt]{4.818pt}{0.400pt}}
\put(220.0,440.0){\rule[-0.200pt]{4.818pt}{0.400pt}}
\put(198,440){\makebox(0,0)[r]{300}}
\put(1416.0,440.0){\rule[-0.200pt]{4.818pt}{0.400pt}}
\put(220.0,550.0){\rule[-0.200pt]{4.818pt}{0.400pt}}
\put(198,550){\makebox(0,0)[r]{400}}
\put(1416.0,550.0){\rule[-0.200pt]{4.818pt}{0.400pt}}
\put(220.0,659.0){\rule[-0.200pt]{4.818pt}{0.400pt}}
\put(198,659){\makebox(0,0)[r]{500}}
\put(1416.0,659.0){\rule[-0.200pt]{4.818pt}{0.400pt}}
\put(220.0,768.0){\rule[-0.200pt]{4.818pt}{0.400pt}}
\put(198,768){\makebox(0,0)[r]{600}}
\put(1416.0,768.0){\rule[-0.200pt]{4.818pt}{0.400pt}}
\put(220.0,877.0){\rule[-0.200pt]{4.818pt}{0.400pt}}
\put(198,877){\makebox(0,0)[r]{700}}
\put(1416.0,877.0){\rule[-0.200pt]{4.818pt}{0.400pt}}
\put(244.0,113.0){\rule[-0.200pt]{0.400pt}{4.818pt}}
\put(244,68){\makebox(0,0){1}}
\put(244.0,857.0){\rule[-0.200pt]{0.400pt}{4.818pt}}
\put(482.0,113.0){\rule[-0.200pt]{0.400pt}{4.818pt}}
\put(482,68){\makebox(0,0){1.2}}
\put(482.0,857.0){\rule[-0.200pt]{0.400pt}{4.818pt}}
\put(721.0,113.0){\rule[-0.200pt]{0.400pt}{4.818pt}}
\put(721,68){\makebox(0,0){1.4}}
\put(721.0,857.0){\rule[-0.200pt]{0.400pt}{4.818pt}}
\put(959.0,113.0){\rule[-0.200pt]{0.400pt}{4.818pt}}
\put(959,68){\makebox(0,0){1.6}}
\put(959.0,857.0){\rule[-0.200pt]{0.400pt}{4.818pt}}
\put(1198.0,113.0){\rule[-0.200pt]{0.400pt}{4.818pt}}
\put(1198,68){\makebox(0,0){1.8}}
\put(1198.0,857.0){\rule[-0.200pt]{0.400pt}{4.818pt}}
\put(1436.0,113.0){\rule[-0.200pt]{0.400pt}{4.818pt}}
\put(1436,68){\makebox(0,0){2}}
\put(1436.0,857.0){\rule[-0.200pt]{0.400pt}{4.818pt}}
\put(220.0,113.0){\rule[-0.200pt]{292.934pt}{0.400pt}}
\put(1436.0,113.0){\rule[-0.200pt]{0.400pt}{184.048pt}}
\put(220.0,877.0){\rule[-0.200pt]{292.934pt}{0.400pt}}
\put(45,495){\makebox(0,0){$m_{3/2}$[GeV]}}
\put(828,23){\makebox(0,0){$\theta$[rad]}}
\put(721,331){\makebox(0,0)[l]{ The Allowed Range}}
\put(220.0,113.0){\rule[-0.200pt]{0.400pt}{184.048pt}}
\put(220,222){\usebox{\plotpoint}}
\multiput(339.00,220.92)(5.590,-0.492){19}{\rule{4.427pt}{0.118pt}}
\multiput(339.00,221.17)(109.811,-11.000){2}{\rule{2.214pt}{0.400pt}}
\put(220.0,222.0){\rule[-0.200pt]{28.667pt}{0.400pt}}
\multiput(1174.00,211.58)(5.590,0.492){19}{\rule{4.427pt}{0.118pt}}
\multiput(1174.00,210.17)(109.811,11.000){2}{\rule{2.214pt}{0.400pt}}
\put(458.0,211.0){\rule[-0.200pt]{172.484pt}{0.400pt}}
\put(1293.0,222.0){\rule[-0.200pt]{28.667pt}{0.400pt}}
\put(220,746){\usebox{\plotpoint}}
\multiput(220,746)(11.186,-17.483){11}{\usebox{\plotpoint}}
\multiput(339,560)(17.492,-11.172){7}{\usebox{\plotpoint}}
\multiput(458,484)(19.787,-6.266){6}{\usebox{\plotpoint}}
\multiput(578,446)(20.241,-4.593){6}{\usebox{\plotpoint}}
\multiput(697,419)(20.667,-1.910){6}{\usebox{\plotpoint}}
\multiput(816,408)(20.729,-1.045){5}{\usebox{\plotpoint}}
\multiput(935,402)(20.730,1.036){6}{\usebox{\plotpoint}}
\multiput(1055,408)(20.570,2.766){6}{\usebox{\plotpoint}}
\multiput(1174,424)(20.001,5.546){6}{\usebox{\plotpoint}}
\multiput(1293,457)(19.520,7.054){6}{\usebox{\plotpoint}}
\put(1412,500){\usebox{\plotpoint}}
\end{picture}
\end{center}
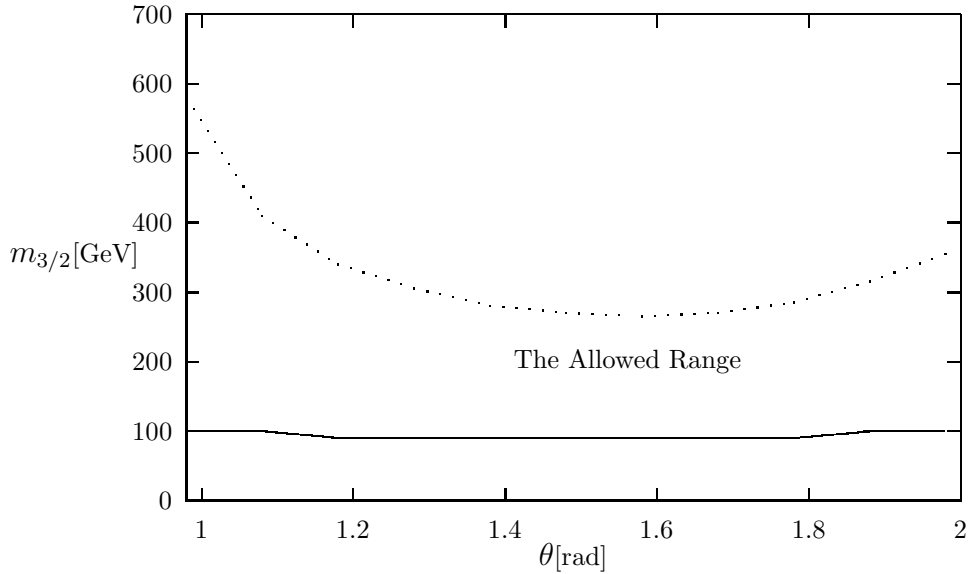
Figure 5: The allowed parameter space with $\mu <0$. The solid 
line corresponds to the constraints from $b\rightarrow s \gamma$ 
while the dashed line corresponds to the upper bounds on $m_{3/2}$ from 
the neutralino relic abundance.
\vspace{0.5cm}

\section{\large{\bf{The LSP detection rates }}}
\hspace{0.75cm}In models with conserved R-parity the lightest supersymmetric 
particle (LSP) is considered the favorite candidate for cold dark matter 
(CDM). As mentioned in section 1, in a previous
analysis\cite{shafi}  we have shown that the lightest neutralino in the
minimal string unification 
turns out to be  the LSP  and it is almost a pure bino. Moreover, 
requiring $ 0.1 \leq \Omega_{LSP} \leq 0.9$, with $ 0.4\leq h
\leq 0.8$, leads to relevant constraints on the parameter 
space (\mg\ , \th\ ). This leads to a stringent upper bound on the LSP 
mass of about 160 GeV in the case of pure dilaton supersymmetry breaking. 
In addition, severe limits on the parameter space were obtained in the 
last section by imposing the constraints that derive from $b \rightarrow 
s \gamma $. In this section we are interested in the
detectability of the LSP of 
this model taking account of all relevant constraints.\\

It was shown in Ref.\cite{borzumati} that the detectability of neutralino 
dark matter is linked to the amplitude for $b \rightarrow s 
\gamma$, and the experimental bounds on the branching ratio for 
the inclusive $b \rightarrow s \gamma$ decay impose strong 
constraints on the region of the parameter space where 
sizable counting rates for relic neutralinos are expected. 
The main reason for this is that both the counting rate and the
branching ratio increase with decreasing mass of the 
Higgs bosons. In Ref.\cite{nath} it was shown that
there are sizable regions of the parameter space with $R >0.01$ including
this constraint. Other authors~\cite{gondolo} have recently claimed that
there exists some 
possibility for 
further enhancement of the detectability of neutralino scattering with
nuclei. It is 
certainly relevant to investigate the impact of the various 
restrictions (including those coming from $b\rightarrow s \gamma$ in 
section 3) on the neutralino-nuclei scattering in the class of 
superstring models under discussion.\\

Perhaps the most natural way of searching for the neutralino dark 
matter is provided by direct experiments, where the 
effects induced in appropriate detectors by neutralino-nucleus 
elastic scattering may be measured. 
The differential detection rate is given by
\be
\frac{d R}{d Q} = \frac{\sigma \rho_{\chi}}{2 m_{\chi} m_r^2} F^2(Q) 
\int_{v_{min}}^{\infty} \frac{ f_1(v)}{v} dv,
\ee
where $f_1(v)$ is the distribution of speeds relative to the detector.
The reduced mass is $m_r= (\frac{Q m_N}{2 m_r^2})^{1\over 2}$, where $
m_N$ is
the mass of the nucleus, $ v_{min}= (\frac{Q m_N}{2 m_r^2})^{1/2}$, $Q$
is the energy deposited in the detector and $\rho_{\chi}$ is the
density of the neutralino near the Earth. It is common to fix
$\rho_{\chi}$ to be 
$\rho_{\chi}=0.3 GeV/cm^3$. Instead, we will determine it  
from the relation
\be
\rho_{\chi} = \Omega_{\chi} h^2 \times \rho_{critical} ,
\ee
where $\rho_{critical} \sim 1.8 \times 10^{-29} g/cm^3$ and $\Omega_{\chi} 
h^2$ is the neutralino relic density, so we are treating $\rho_{\chi}$ as 
a function of the neutralino mass. We will compare the result with the 
one we would obtain if $\rho_{\chi}=0.3 GeV/cm^3$. The quantity $\sigma$
is the elastic-scattering cross
section of the LSP with a given nucleus. In 
general $\sigma$ has two contributions: spin-dependent contribution 
arising from $Z$ and $\tilde{q}$ exchange diagrams, and spin-independent 
(scalar) contribution due to the Higgs and squark exchange diagrams. For 
$^{76}Ge$ detector, where the total spin of $^{76}Ge$ is equal to zero, 
we have contributions only from the scalar part. The form factor in this 
case is given by~\cite{engel}
\be
F(Q) = \frac{3 j_1(q R_1) }{ q R_1} e^{-\frac{1}{2} q^2 s^2},
\ee
where the momentum transferred is $q= \sqrt{2 m_N Q}$, $R_1=(R^2 -5 
s^2)^{1/2}$ with $R=1.2 fm A^{1/2}$ and A is the mass number of $^{76}Ge$. 
$j_1 $ is the spherical Bessel function and $s \simeq 1 fm $.\\

	The event rate $R$ is  presented in 
Figs.6 and 7 for the two cases, $\rho_{\chi}$ as function of $
m_{\chi}$,
and $\rho_{\chi}=0.3 GeV/cm^3$. 
The detection rates are of order $10^{-1} - 10^{-4}$ events/kg/day. Also,
we can see that the result significantly changes when we treat
$\rho_{\chi}$ as a function of the
neutralino mass.
\vspace{0.5cm}
\begin{center}
\input dir1.tex
\end{center}
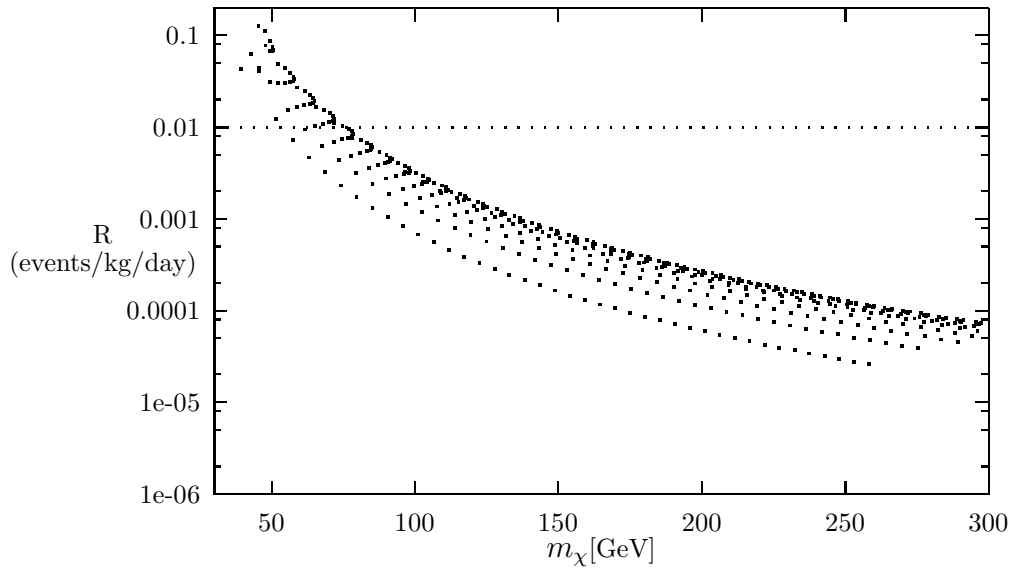
Figure 6: The event rate $R$ versus $m_{\chi}$ with
$\rho_{\chi}$ treated as function of $m_{\chi}$, and $\tan \beta \simeq
2$. The horizontal line denotes the present experimental sensitivity. 
\vspace{0.5cm}
\begin{center}
\input dir2.tex
\end{center}
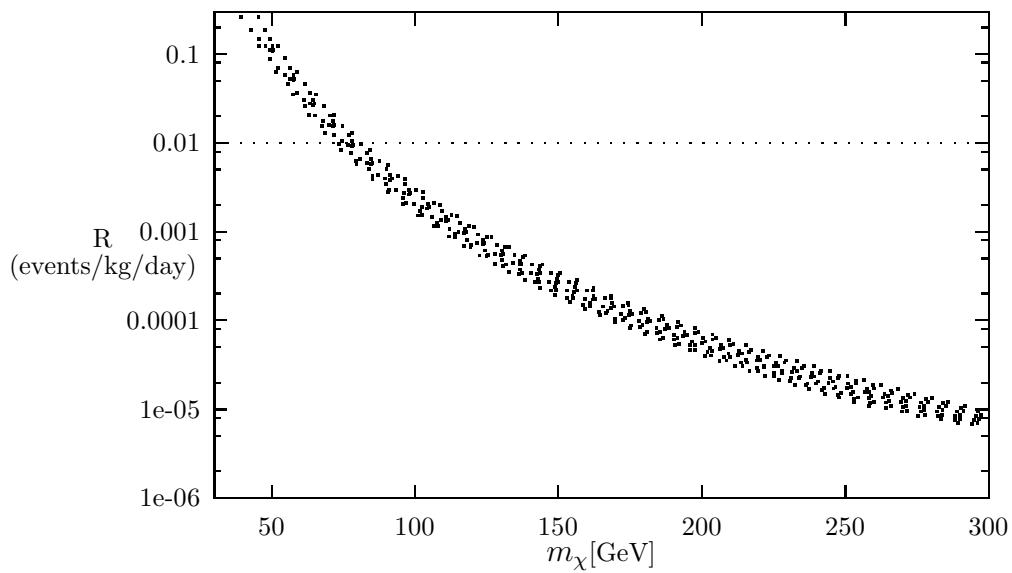
Figure 7: The event rate $R$ versus $m_{\chi}$
for $\rho_{\chi}=0.3 GeV/cm^3$  and $\tan \beta \simeq 2$. The horizontal
line as in Fig.6.  
\vspace{0.5cm}
	
	A promising method for indirect detection of neutralinos 
in the halo is the observation of the energetic neutrinos from 
the annihilation of neutralinos that accumulate in the sun or in the
earth. Among the annihilation products are ordinary neutrinos which may
be observable in suitable detectors. The energies of the neutrinos are
about
a third of the LSP mass so they are easily distinguished from solar
neutrinos or any other known background. 
The technique for the detection of such energetic neutrinos is through 
observation of upward muons produced by the charged current interactions
of the neutrinos in the rock below the detector. Concentrating on the
neutralino annihilation on the sun, the flux of such muons
from neutralino annihilation can be written as
\be
\Gamma = 2.9 \times 10^8 m^{-2} yr^{-1} \tanh^2(t/\tau) \rho_{\chi}^{0.3} 
f(m_{\chi}) \zeta(m_{\chi}) (\frac{m_{\chi}^2}{GeV})^2 
(\frac{f_P}{GeV^{-2}})^2 .
\ee
The neutralino-mass dependence of the capture rates is described
by~\cite{report}
\be
f(m_{\chi}) = \sum_i f_i \phi_i S_i(m_{\chi}) F_i(m_{\chi}) \frac{m_i^3 
m_{\chi}}{(m_{\chi}+m_i)^2},
\ee
where the quantities $\phi_i$ and $f_i$ describe the distribution of
element 
$i$ in the sun and they are listed in Ref.~\cite{report}, the quantity 
$S_i(m_{\chi})=S(\frac{m_{\chi}}{m_{N_i}})$ is the kinematic suppression 
factor for capture of neutralino of mass $ m_{\chi}$ from a nucleus of 
mass $m_{N_i}$~\cite{report} and $F_i(m_{\chi})$ The form factor
suppression for the capture of a 
neutralino of mass $ m_{\chi}$ by a nucleus $i$. Finally, the function
$\zeta (m_{\chi})$ describes the energy 
spectrum from neutralino annihilation for a given mass.\\
	
	In Figs.8 and 9 we present the results for muonic fluxes 
resulting from captured neutralinos in the sun for $\rho$ as a function 
of the neutralino mass, and for $\rho_{\chi}=0.3 GeV/cm^3$.
We see that the predicted muonic flux lies between $10^{-2}$ and
$10^{-9}$ muon/$m^2$/day. Clearly, large scale detectors are best suited
for neutralino detection.
\vspace{0.5cm}
\begin{center}
\input in1.tex
\end{center}
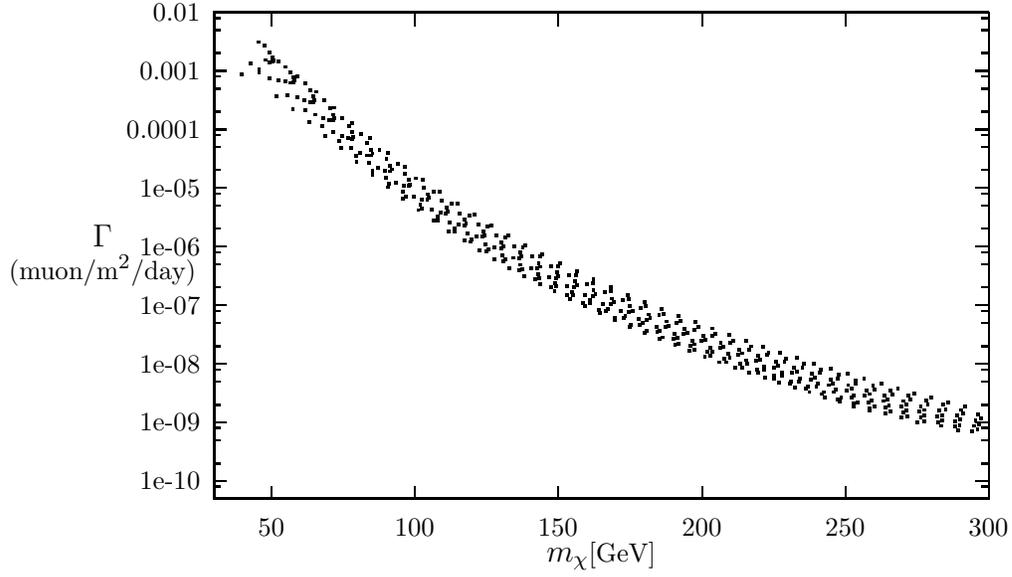
Figure 8: The mounic flux $\Gamma$ versus $m_{\chi}$
with $\rho_{\chi}$ considered a function of $m_{\chi}$, and $\tan \beta
\simeq 2$.
\vspace{0.5cm}
\begin{center}
\input in2.tex
\end{center}
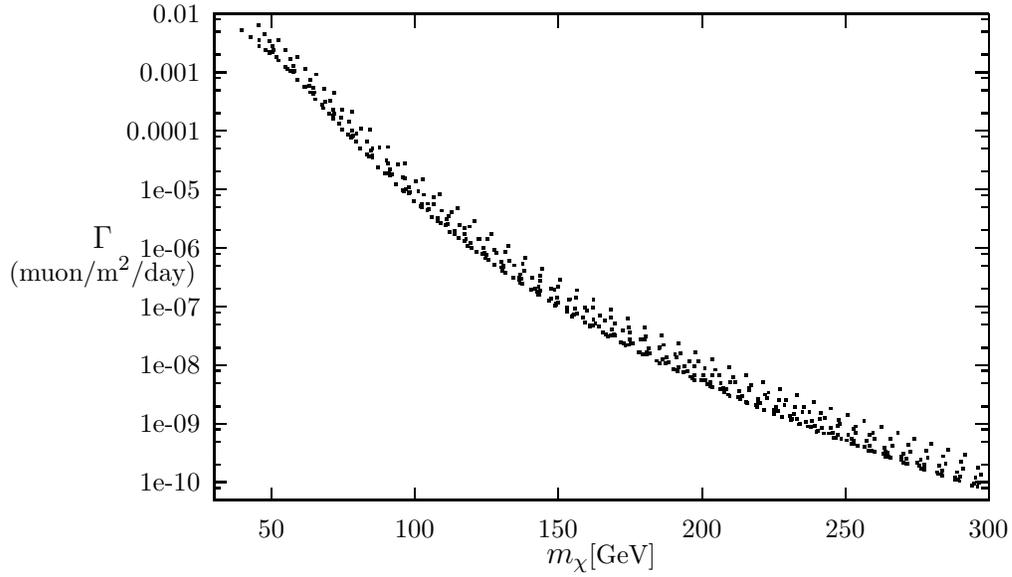
Figure 9: The mounic flux $\Gamma$ versus $m_{\chi}$
for $\rho_{\chi}=0.3$ and $\tan \beta \simeq 2$.
\vspace{0.5cm}
\newpage
\section{Conclusions}

\hspace{0.75cm}The decay $b\rightarrow s \gamma$ has been employed to
derive the most stringent lower bound on the gravitino mass in the
`minimal' string unification models. This leads to lower bounds on the
sparticles and Higgs mass spectra. By combining this information with the
upper bounds available from considerations of the LSP (`bino') relic
abundance, we are able to estimate the direct and indirect detection rates
for the latter. Large scale detectors are needed to discover the LSP of
our scheme.\\

\noindent{\Large\bf Acknowledgments}
\vskip0.5truecm
 The authors would like to thank G.Jungman for useful discussion. 
S.K would like to acknowledge the hospitality of ICTP. Q.S and A.M 
acknowledge 
support by the US Department of Energy, Grant No. DE-FG02-91ER and European
TRM contract ERBFMRXCT 960090, respectively.


\begin{thebibliography}{99}

\bibitem{nilles}
For a review, see for instance:
H.P. Nilles, \prep{110}{84}{1}; A.B. Lahanas and D.V. Nanopoulos,
\prep{145}{87}{1};
H.E. Haber and G. Kane, \prep{117}{85}{75};
 
\bibitem{calabi}
P.Candelas, G.Horowitz, A.Strominger and E.Witten, \npb{258}{85}{46}

\bibitem{orbifold}
L.Dixon, J.Harvey, C.Vafa and E.Witten, \npb{261}{85}{678}; 
\npb{274}{86}{285}
L.E.Iba\~{n}ez, J.Mas, H.P.Nilles and F.Quevedo, \npb{301}{88}{157}

\bibitem{ibanez1}
L.E. Iba\~{n}ez and D. L\"{u}st, \npb{382}{92}{305};

\bibitem{kap}
V.S. Kaplunovsky and L. Louis, \plb{306}{93}{269}. 

\bibitem{munoz}
B. de Carlos, J.A. Casas and C. Mu\~{n}oz, \npb{399}{93}{623} and
\plb{299}{93}{234}.
 
\bibitem{ibanez2}
A. Brignole, L.E. Iba\~{n}ez, C. Mu\~{n}oz,
\npb{422}{94}{125};
Erratum, \ibid{436}{95}{747}.
 
\bibitem{kobay}
T.Kobayashi, D.Suematsu, K.Yamada and Y.Yamagishi, \plb{348}{95}{402}; 
A. Brignole, L.E. Iba\~{n}ez, C. Mu\~{n}oz and C.Scheich, {\bf 
hep-ph/9508258}; Y.Kawamura, S.Khalil and T.Kobayashi, {\bf
hep-ph/9703239}.

\bibitem{ross}
L. Iba\~{n}ez, D. Lust and G.G. Ross, \plb{272}{91}{261};
L.E. Iba\~{n}ez and D. Lust, \npb{382}{92}{305}. See also J.Ellis, S.Kelley
and D.V.Nanopoulos, \plb{272}{91}{31}; M.Cvetic, Proceedings of Dallas HEP
1992, 1178.

\bibitem{vissani}
S.Khalil, A.Masiero and F.Vissani, \plb{375}{96}{154}.

\bibitem{shafi} 
S.Khalil, A.Masiero and Q.Shafi, {\bf hep-ph/9611280}, to appear in Phys.
Lett.B.

\bibitem{witten}
E.Witten \plb{155}{85}{151};
S.Ferrara, C.Kounnas and M.Porrati, \plb{181}{86}{99}
M. Cveti\v{c}, J.Louis and B.Ovrut \plb{206}{88}{99}
M. Cveti\v{c}, J.Molera and B.Ovrut \prd{40}{89}{1140}

\bibitem{masiero1} 
G.F. Giudice and A. Masiero, \plb{206}{88}{480};

\bibitem{kap2}
V.S.Kaplunovsky, \npb{307}{88}{145}.

\bibitem {gamberini}
G.Gamberini, G.Ridolfi and F.Zwirner \npb{331}{90}{331}.

\bibitem{savoy}
J.P.Derendinger and C.A.Savoy, \npb{237}{84}{307}.

\bibitem{amer}
R. Ammar et a. (CLEO collaboration), Phys.
Rev.Lett. 71(1993) 674; B. Barish et al.(CLEO collaboration)  

\bibitem{masiero2}
S. Bertolini, F. Borzumati, A. Masiero and
G. Ridolfi, \npb{353}{91}{591}.

\bibitem{bertolini}
R.Barbieri and G.F. Giudice, \plb{309}{93}{86}
N.Osimo, \npb{404}{93}{20};
R.Garisto and J.N. Ng \plb{315}{93}{372};
Y.Okad, \plb{315}{93}{119};
M.Diaz \plb{322}{94}{207};
F.Borzumati, \zpc{63}{94}{395};
J.L. $\rm{L\acute{o}pez}$, D.V. Nanopoulos, G.T. Park and A. Zichichi,
\prd{49}{94}{355};
R. Arnowitt and P.Nath, \plb{336}{94}{395}; 
S.Bertolini and F. Vissani, \zpc{67}{95}{513};
B.de Carlos, J.A. Casas, \plb{349}{95}{300}.
 
\bibitem{karniotis}
G.Kraniotis, \zpc{71}{96}{163};
B. de Carlos and G.V. Kraniotis {\bf hep-ph/9610355}.\\
For a discussion in the context of the no-scale supergravity 
$SU(5)\times U(1)$  see J.Lopez,D.V. Nanopoulos et 
al, \prd{51}{95}{147}.

\bibitem{misiak}
S.Bertolini, F.Borzomati and A.Masiero, \prl{59}{87}{180};
N.G. Deshpande, P. Lo, J. Trampetic, G. Eilam, and P. Singer,
\prl{59}{87}{183};
B. Grinstein, R. Springer, and M .Wise \plb{202}{88}{138};
R. Grigjanis et al \plb{213}{88}{355};
M. Misiak \plb{269}{91}{161};
M. Misiak , \npb{393}{93}{23};
M.Ciuchini, E.Franco, G.Martinelli and L.Silvestrini \plb{316}{93}{127};
G.Cella, G.Curci, G.Ricciardi and A.Vicere \plb{325}{94}{227}.
\bibitem{guduice}
R.Barbieri and G.F. Giudice, in ref.~\cite{bertolini}

\bibitem{ng}
R. Garisto and J.N. Ng in ref.~\cite{bertolini};

\bibitem{borzumati}
F.M Borzumati, M. Drees and M.M. Nojiri,\prd{51}{95}{341} ; 

\bibitem{nath}
R. Arnowitt and P.Nath, \prd{54}{96}{2374};

\bibitem{gondolo}
L.Bergstrom, P.Gondolo, Astropart.Phys.5:263-278,1996.

\bibitem{engel}
J.Engel \plb{264}{91}{114}.
\bibitem{report}
See G.Jungman, M.Kamionkowski and K.Griest, \prep{267}{96}{195}, and
 
\end{thebibliography}
\end{document}